# OCTAD-S: Digital Fast Fourier Transform Spectrometers by FPGA

Kazumasa Iwai[*,1,2], Yûki Kubo[2], Hiromitsu Ishibashi[2], Takahiro Naoi[2], Kenichi Harada[3], Kenji Ema[3], Yoshinori Hayashi[3], and Yuichi Chikahiro[3]

1. Institute for Space-Earth Environmental Research, Nagoya University, Furo-cho, Chikusa-ku, Nagoya, 464-8601, Japan; k.iwai@isee.nagoya-u.ac.jp
2. National Institute of Information and Communications Technology, 4-2-1 Nukui-kita, Koganei, Tokyo 184-8795, Japan
3. Engineering Department, ELECS Industry CO. LTD, 1-22-23 Takatsu, Kawasaki, Kanagawa 213-0014, Japan

Abstract

We have developed a digital fast Fourier transform (FFT) spectrometer made of an analog-to-digital converter (ADC) and a field-programmable gate array (FPGA). The base instrument has independent ADC and FPGA modules, which allow us to implement different spectrometers in a relatively easy manner. Two types of spectrometers have been instrumented, one with 4.096 GS/s sampling speed and 2048 frequency channels and the other with 2.048 GS/s sampling speed and 32768 frequency channels. The signal processing in these spectrometers has no dead time and the accumulated spectra are recorded in external media every 8 ms. A direct sampling spectroscopy up to 8 GHz is achieved by a microwave track-and-hold circuit, which can reduce the analog receiver in front of the spectrometer. Highly stable spectroscopy with a wide dynamic range was demonstrated in a series of laboratory experiments and test observations of solar radio bursts.

## Introduction

Spectroscopy has been an important measurement method for science since a long time. It has been widely used in astronomy, space science, and geophysical sciences; e.g., the solar and planetary radio bursts observed in the deci- to deca-meter range (e.g., McLean and Labrum 1985), and the molecular lines from the astronomical object mainly in the microwave to far-infrared wavelength range using the heterodyne technique (e.g., Weinreb et al. 1963).

As the needs for spectroscopy have increased, spectroscopy techniques and spectrometers have been improved. Radio spectroscopy has enabled a filter bank technique that employs an array of analog bandpass filters. The number of frequency channels of their spectra is limited by the number of bandpass filters. Frequency-sweep or frequency-agile type spectrum analyzers, in which the frequency of the local oscillator is swept or changed sequentially, have also been commonly used, while usually the sensitivities of their spectra are limited because the accumulation time of each frequency channel is much shorter than the sweep time. Acousto-optical spectrometers (AOS), which utilize diffraction of light, have been used mainly for heterodyne observations in the high-frequency range. However, this technique has limitations of dynamic range and stability.

Digital technology has dramatically improved radio spectroscopy and many digital spectrometers have been developed. In the digital spectrometer, the analog signal is digitized by analog-to-digital converters (ADCs) and converted to spectra by digital signal processing, which provides stable spectroscopy. Traditional digital spectrometers were of an autocorrelation-type in which the digitized signal was first autocorrelated by a customer specific chip and then Fourier transformed (Belgacem et al. 2004). Recently, fast Fourier transform (FFT) type spectrometers have become more common. In this type of spectrometer, the digitized signal is first fast Fourier transformed, and then, it is squared to form a power spectrum. This FFT type digital spectroscopy can reduce the amount of necessary calculations, especially for a spectrum with a large number of frequency channels. Several digital FFT spectrometers have been developed for radio astronomy (e.g., Benz et al. 2005, Klein et al. 2006; 2012, Kamazaki et al 2012). These spectrometers are used in various radio telescopes (e.g., Heyminck et al. 2012, Iwai et al. 2012).

One of the important merits of using a field-programmable gate array (FPGA) is that the digital circuit can be rewritten easily. Therefore, we can use the same device for different tasks by modifying the signal processing circuit in the FPGA. However, the different functions usually require not only the modification of the signal processing program but also different specifications for the hardware. For example, if we want to increase the number of FFT points, we usually have to use a better FPGA processor. The flexibility of our hardware design, which allows us to modify the ADC and FPGA configuration, would reduce the total cost of the spectrometer.

The sampling speed of the current ADCs is not enough to sample the entire observation band of typical radio telescopes. Hence, in many telescopes, a radio signal is frequency converted to the appropriate frequency range for the digital spectrometers. Hence, digital spectrometers require a more complicated backend receiver system than the sweep-type spectrum analyzers. A frequency that is higher than the Nyquist frequency of the ADCs will appear as an aliasing signal. The over-Nyquist frequency signal can be sampled by using a better track-and-hold circuit in front of ADCs, which allows the instrument to make spectra of higher frequency signals than the Nyquist frequency of the ADCs.

Recently, the demand for wideband and stable spectroscopy with a wide dynamic range has increased. For example, solar radio bursts with fine spectral structures were observed in a wide frequency range (e.g. Dąbrowski et al 2011 and references therein). In addition, various telescopes are planned to conduct simultaneous multi-molecular-line observations in millimeter and sub-millimeter wavelength range (e.g. Minamidani et al, 2016).

In this paper, we report the development of digital FFT spectrometers and the results of their evaluation tests. Our spectrometer has independent FPGA and ADC modules, which allow us to change these devices easily. Our instruments have a track-and-hold circuit that accepts microwave input, which enables direct sampling spectroscopy up to 8 GHz using an ADC with a 4.096 GS/s sampling speed. This instrument can provide stable and wideband spectra with a wide dynamic range for various spectroscopic studies. The following sections describe the hardware and signal processing of our spectrometer, results of performance evaluation tests, and some observational results.

## Method and Materials

### Hardware: OCTAD-S

The spectrometers are developed based on a digital instrument platform named OCTAD (Optically Connected Transmission system for Analog to Digital Conversion: Figure 1), which is designed and manufactured by ELECS Industry Co. Ltd[1]. The basic concept behind this general-purpose digital instrument is the combination of ADCs and FPGAs. This instrument is composed of three main modules: the ADC module; a digital signal processing (DSP) module, which contains the FPGAs; and a main module, which contains the local oscillator, the data interface to external media, etc. The ADC and DSP modules are independent of the main module, which allows us to upgrade the ADC and FPGA devices at a lower cost. We developed digital spectrometers using OCTAD by coding the FFT processing on the FPGA and named the resulting instrument OCTAD-S (S: Spectrometer).

---

[1] http://www.elecs.co.jp/

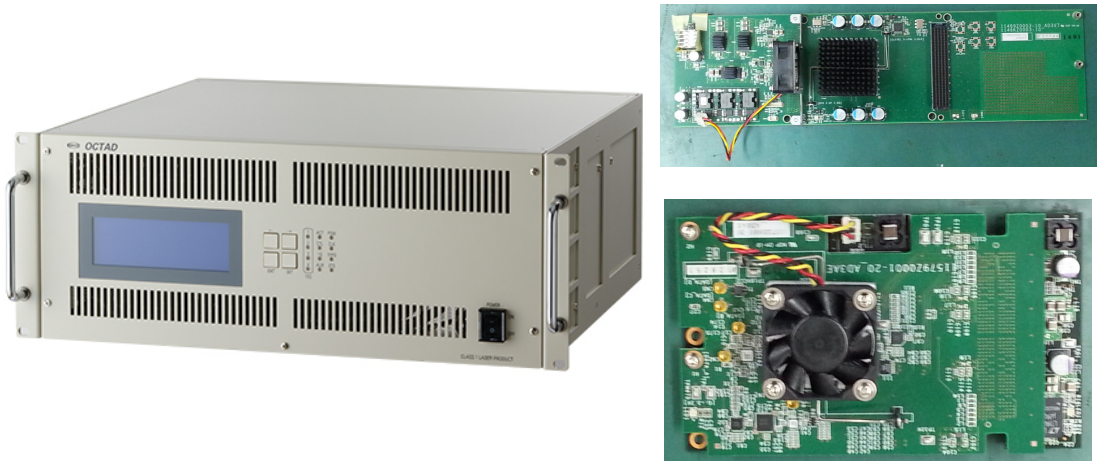

Figure 1. Photographs of the OCTAD-S. (Left) Overview of the instrument. (Upper right) DSP module and (lower right) ADC module.

The instrument is synchronized by an external 10 MHz reference clock. It also has an internal phase-locked oscillator (PLO). Hence, the instrument works, by using PLO, even when external clocks are unavailable. The ADC module has a track-and-hold circuit (HMC661LC4B) that accepts microwave input, and a 10-bit ADC (EV10AQ190A), which realizes a maximum sampling speed of 5 GS/s by interleaving four 1.25 GS/s ADC cores with a typical effective number of bits (ENOB) of 8 bit. The track-and-hold circuit is implemented in front of the ADC, and it converts the single-ended input signal to a differential signal. A phase shifter is also implemented to connect the track-and-hold circuit and ADC in order to adjust the phases between the two components. Consequently, the spectrometer realizes direct sampling spectroscopy from 0 to 8 GHz although the ADC of OCTAD-S does not have a good response in the microwave range (see the next section). The DSP module has a FPGA chip, Xilinx Virtex-7 (XC7VX485T), which has 485,760 logic cells. The interface part of this instrument transfers the processed data, which are power spectra for the spectrometers, to external media. It has several options: G-bit Ethernet, 10 G-bit Ethernet, and 10 G-bit optical transceiver. These broadband data transfer options enable us to transfer massive amounts of data quickly. The front panel of this instrument contains some LEDs that are used to check the conditions of the instrument. For example, we can check the conditions of the external clock or the overflow of the ADCs. There is also a liquid crystal display in the front panel that shows the typical ADC count levels in real time.

*Signal processing*

Figure 2 shows the block diagram of the signal processes in FPGA. In this instrument, FPGA processing is performed in the integer format, implemented as IP cores provided by Xilinx. The ADC outputs are decoded linearly and multiplied by a window function. There are four options for the window function (none, Hanning, Blackman, and custom). The custom option accepts user-defined window functions, which can be uploaded to the instrument via FTP. Here, we use the provided Blackman function. Polyphase filterbank (PFB) based digital spectroscopy is known to provide a better performance than the window function based FFT spectroscopy (e.g., Price 2016). The PFB based spectroscopy places polyphase finite-impulse-response (FIR) filters in front of the FFT processing component instead of a window function, which requires a bit more signal processing resources than the window function based FFT. Therefore, we used a window function in order to save the signal processing resources. The difference between the window function based FFT and PFB will be discussed in the next section.

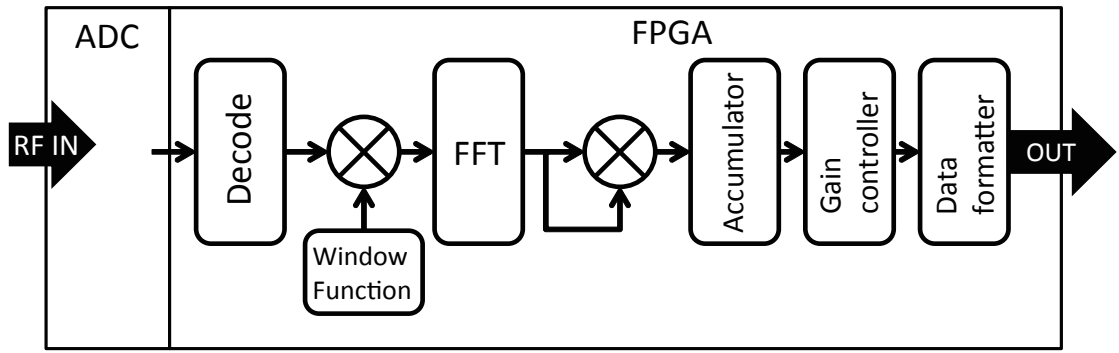

Figure 2. Block diagram of the signal processes from ADC to FPGA.

The power spectrum, which has a 32-bit resolution, is generated by the FFT IP core (LogiCORE IP Fast Fourier Transform v9.0), which realizes pipeline processing. The throughput of this FFT IP core measured by our FPGA clock is smaller than the sampling rate of the ADC. Therefore, we implemented several FFT IP cores in a FPGA, and we operate them in parallel. A buffer memory, implemented before the FFT to store the data, can prevent the data loss. These implementations enable the instrument to make real-time FFT spectra with no dead time. The power spectra derived from each of the FFT IP cores are accumulated in a memory (called accumulator) and sent to the next step at every accumulation time. The accumulation time depends on the bit length of the accumulator, which is decided by the user from between 40 to 48 bits, and the corresponding accumulation time is from 0.0008 s (8 ms) to 2 s. We set it to 8 ms by using a 40 and 45 bit accumulator for 2G64K and 4GK, respectively.

To reduce the amount of data, the gain controller, which receives the accumulated spectra, extracts only a 16-bit fraction from the 40–45-bit spectrum. The range of the extracted 16-bit, which we call the FFT gain, can be defined by the user with Telnet, where the user does not have to change the FPGA logic to set the FFT gain. This option can be also used to extend the accumulation time. The extracted 16-bit spectrum is formatted and saved to the external media via G-bit Ethernet within 8 ms. Each spectrum has its own header that includes a timestamp of the hardware. The instrument should be synchronized by an external 10 MHz reference signal that has a precisely correct time to enable scanning observations. The ADC outputs an overflow bit, which is also included in the header, and can be used to check the analog signal level.

The above processes are all coded in FPGA, which means that the spectrometer can make and save spectra independently. Each spectrometer can have its own IP address. We can control the instrument by sending a simple command (e.g., start/stop) via Telnet. Telnet is also used to send some commands to set the window function, accumulation time, and FFT gain in the FPGA.

To meet the diversified needs of spectroscopy, we implemented two spectrometers on OCTAD-S as follows:

> **OCTAD-S 4G4K**: This instrument has four analog input channels. Each channel has a 10-bit ADC, operated at a 4.096 GS/s sampling speed by interleaving four 1.024 GS/s digitalized signals. Each channel has one FPGA chip. The FPGA performs a 4096 point FFT, which yields 2048 frequency channels in the power spectrum. One power spectrum requires 1 μs (= 4,096 point /4.096 GS/s). All signal processes are completed within this time in the FPGA, which means that the instrument has no dead time.

**OCTAD-S 2G64K**: This instrument has two analog input channels. Each channel has a 10-bit ADC, operated at a 2.048 GS/s sampling speed by interleaving four 0.512 GS/s digitalized signals. Each channel has two FPGA chips to achieve faster processing speed using the cascading technique, in which each ADC has two FPGAs operated simultaneously to perform an independent FFT and spectral accumulation in parallel. Thus, accumulated power spectra are derived from the two FPGAs alternately. The FPGAs perform a 65,536 point FFT, which yields 32,768 frequency channels in the power spectrum. One power spectrum requires 32 μs (= 65,536 point /2.048 GS/s). As before, all signal processes are completed within this time in the FPGA, which means that the instrument has no dead time.

Results and Discussion

*Laboratory experiments*

The performances of the two types of our spectrometers were evaluated in laboratory experiments. This section summarizes the results of the experiments.

Figure 3 shows the response of a narrow band input signal generated by a signal generator at 420 MHz. The linearity between the input and output signals are recognized within ~80 dB. It is wider than the dynamic range of 10 bits (~60 dB). This is because the total power of the noise is divided by each frequency channel and refused, while the power of the narrow band input signal is concentrated in one channel.

The spurious free dynamic range (SFDR), defined as the ratio between the input signal and the largest spurious in the spectrum, is usually referred to as one of the characteristics of the spectrometers. However, OCTAD-S uses a time-interleaved ADC containing four separate ADCs, which leads to other types of spurious signals. The gain, offset, and clock phase of the four ADCs are not always perfectly matched. Therefore, a digital spectrometer that uses this kind of ADC creates internal ghost signals especially at the Nyquist frequency ($f_N$) and at $f_N/2$. The interactions between the internal ghosts and the input signal ($f_R$) also become ghosts and appear in the spectra, for example at $f_N/2 \pm f_R$ and $f_N - f_R$. The ghost free dynamic range (GFDR) is defined as the ratio between the input signal and the largest ghost related signal in the spectrum. The blue squares in Figure 3 show the output power of the largest ghost signal at 1444 MHz (420 + 1024). The GFDR of this spectrometer is ~37 dB, which is much larger than the SFDR of ADC (> 50 dB). We can reduce the ghost signals by calibrating the offset, gain, and phase of each interleaving ACD core. When the output of typical ghost spectra, e.g., at $f_N/2 \pm f_R$ and $f_N - f_R$, becomes larger than the output of the input signal of interest, the users can calibrate the interleaving ADCs manually by introducing a continuous wave signal. An automatic calibration option will be an important function for future development.

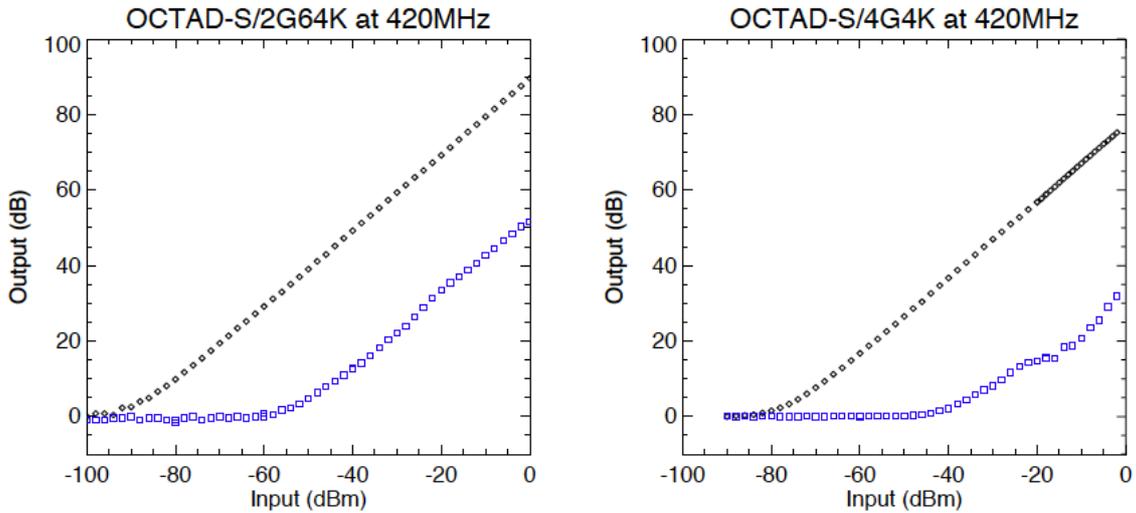

Figure 3. Response of the narrowband input generated by a signal generator at 420 MHz; (left) 2G64K, (right) 4G4K. Black: 420 MHz. Blue: the largest ghost at (left) 604 MHz and (right) 1444 MHz.

In digital spectrometers, a wave that has a frequency higher than the Nyquist frequency of the ADC shows an aliasing effect. An input signal ($f_R$) that is higher than the Nyquist frequency is turned down at the Nyquist frequency and the aliasing signal ($f_a$) can be measured at

$$f_a = -nf_N + f_R \quad n = 0, 2, 4 \ldots$$

$$f_a = (n+1)f_N - f_R \quad n = 1, 3, 5 \ldots$$

Figure 4 shows the response of the narrow band input signal at 420, 3676, 4516, and 7772 MHz. These four input frequencies appear at 420 MHz in OCTAD-S (4G4K). In Figure 4, similar responses are derived from all input frequencies, indicating that the spectrometer realizes direct sampling spectroscopy up to 8 GHz. Digital spectrometers usually require an analog heterodyne receiver system to adjust the input signal frequency. On the other hand, OCTAD-S does not need such a heterodyne receiver system to receive between 0 and 8 GHz. It only requires a band pass filter with an appropriate cutoff frequency. For example, five OCTAD-Ss (4G4K) realize 0–10.24 GHz spectroscopy using only five bandpass filters with cutoff frequencies 0–2.048, 2.048–4.096, 4.096–6.144, 6.144–8.192, and 8.102–10.24 GHz, respectively.

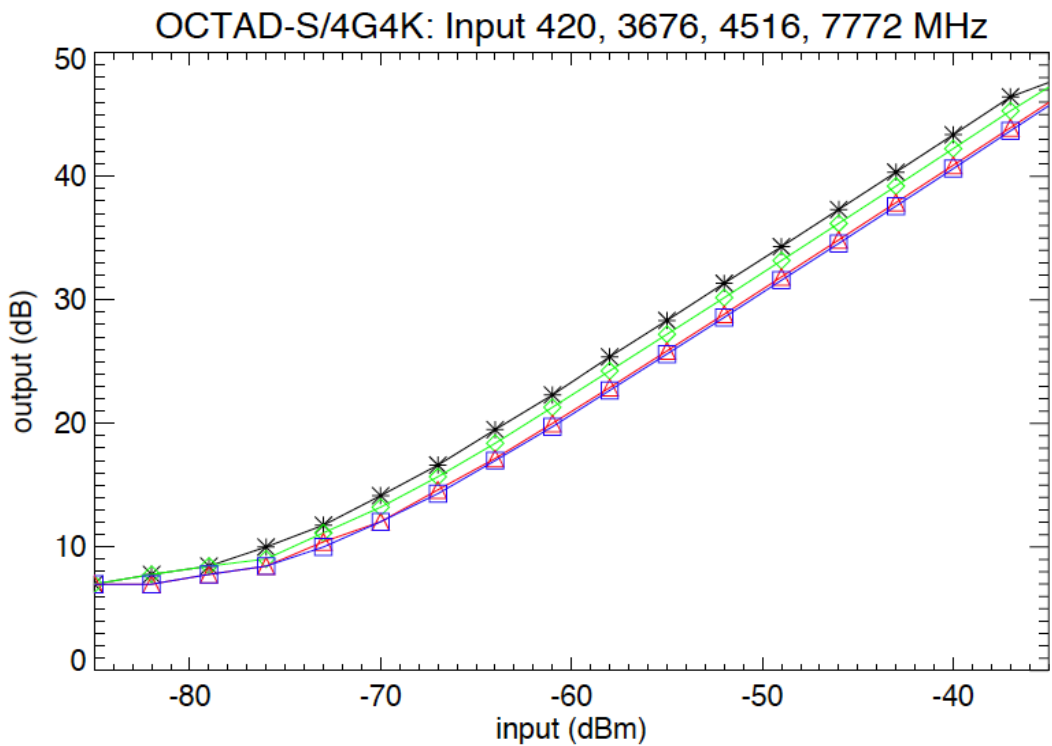

Figure 4. Response of narrow band input at 420 (asterisks), 3676 (diamonds), 4516 (triangles), and 7772 (rectangles) MHz of OCTAD-S 4G4K. Only one FFT gain range (16-bit dynamic range) is shown.

The separation of neighboring frequency channels (channel separation) of our spectrometers is 1 MHz and 31.25 kHz for 4G4K and 2G64K, respectively. However,

FFT spectrometers require a window function before the FFT processing. Hence, the actual spectral resolution should be wider than the channel separation. The equivalent noise bandwidth (ENBW), which is defined as the bandwidth of a rectangular window function whose total passband power is equal to that of the actual window function, is 1.73 for the Blackman function (Harris 1978). The actual spectral resolution was investigated by a filter curve, derived by measuring a fixed frequency channel and, in parallel, producing a narrow band signal and stepping its frequency though the measuring frequency. Figure 5 shows the filter curve derived around 420 MHz. The −10 dB bandwidths are 2.9 MHz and 90 kHz for 4G4K and 2G64K, respectively. The sidelobe level of the filter curve is about −60 dB for both instruments.

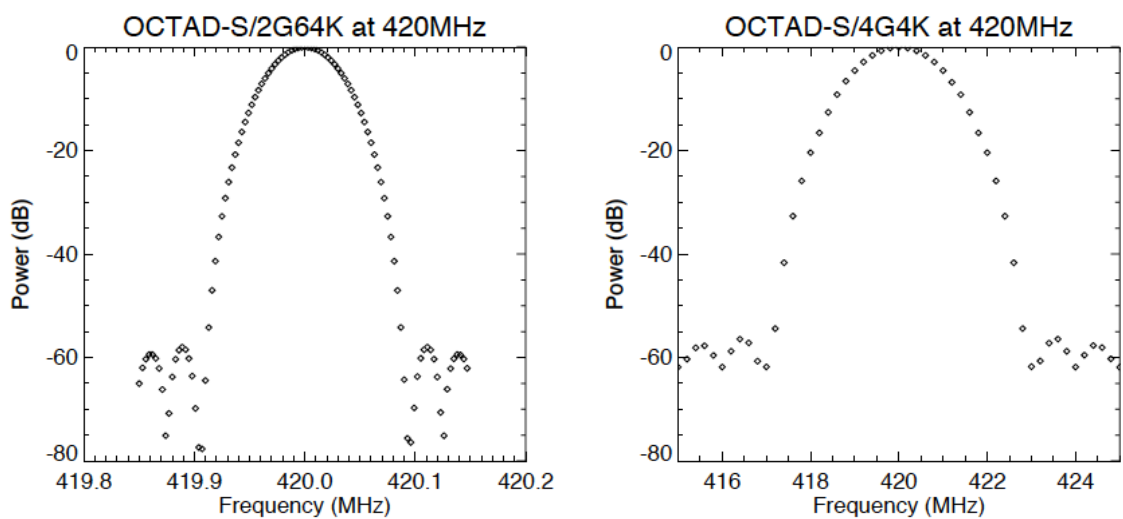

Figure 5. Filter curve at 420 MHz derived from the Blackman window function; (left) 2G64K, (right) 4G4K.

Spectral leakage is defined as a signal that leaks from neighboring channels, which is important for a strong narrowband input such as artificial radio frequency interferences (RFIs). In Figure 5, for example, the leakage of the neighboring channels is about −5 dB, and that of the next neighboring channels is about −20 dB in 4G4K. Therefore, RFI, which is 20 dB stronger than the observed radio signal, can contaminate two neighboring channels; this implies that at least five channels are contaminated by a single narrowband RFI. The shape of the filter curve also affects the derived spectrum of narrowband signals at the edge of the channel, which is defined as scalloping loss. In Figure 5, the scalloping loss is 2–3 dB for both types of spectrometers, which means if a frequency of the narrowband signal is between the channels, a 2–3 dB smaller power spectrum density will be derived at the two neighboring channels. Note that the scalloping loss is not a significant problem if the input signal is much broader than the channel separation of the spectrometer.

PFB spectroscopy can provide a smaller sidelobe level and smaller spectral leakage compared to the window function FFT. The PFB also minimizes the scalloping loss (Price 2016). As mentioned above, we implemented a window function in front of FFT rather than the PFB architecture. Although it is useful in reducing the signal processing resources for this part of the procedure, an implementation of the PFB option can be considered to derive better spectra for future development.

The stability of the spectrometer is defined by the Allan variance. It was measured using a noise source and amplifier in a room whose temperature was controlled by an air conditioner. Figure 6 shows the Allan variance of OCTAD-S 4G4K. The Allan time is longer than 1500 s indicating that the spectrometer has a high stability. Note that this Allan variance also includes the instability of the noise source, amplifier, and room temperature. The room temperature affects the stability of the spectrometer, noise source, and amplifier, which makes the derived Allan time shorter than the actual Allan time of the instrument. In fact, we also measured the Allan time of 2G64K, which is nearly the same as that of 4G4K. However, the Allan time must depend on the bandwidth where the system with a wider bandwidth is supposed to exhibit longer stability time. Therefore, we assumed that the derived Allan time should include the stability of the environments, and the actual Allan time of two types of spectrometers should be better than those derived in this test.

The quality and stability of the spectrum is dependent on the operating conditions, especially the environmental temperature. The most temperature-dependent device included in OCTAD-S is the ADC. On the other hand, the digitized signal is less affected by the temperature. Therefore, the operating temperature range of the ADC limits the operating temperature range of the entire instruments, which is 0–90 °C. Even though the instrument is operated in the appropriate temperature range, the change in temperature during operation affects the ADC output. Therefore, a rapid temperature variation should be avoided to achieve a stable and high-quality continuous spectroscopy. A temperature compensation technique can reduce the temperature variation effect and improve the stability of the spectra. For example, an analog amplifier with an inverse temperature response from that of the ADC can be installed in front of the ADC module to cancel the temperature variation. We can also use the FPGA for a digital temperature compensation. If the temperature response of ADC shows repeatability, it can be measured in the laboratory. Then, a real-time digital temperature compensation using the FPGA and an additional thermometer would be possible by implementing a temperature compensation circuit on the FPGA.

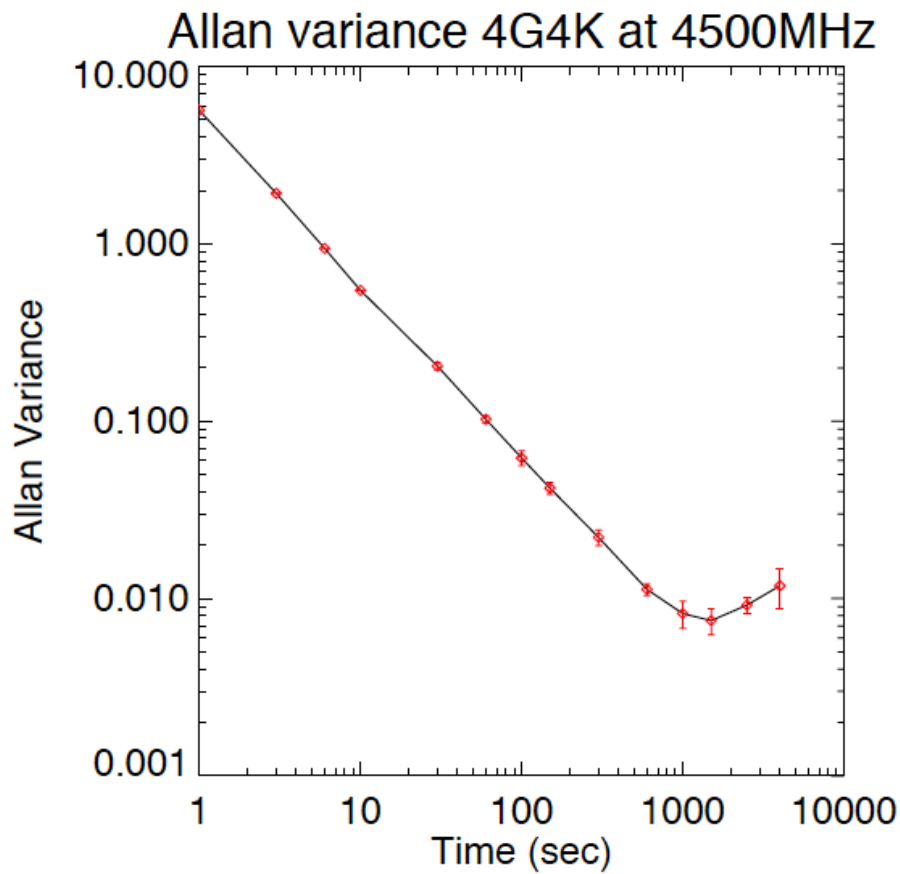

Figure 6. Allan variance of OCTAD-S 4G4K at 4500 MHz.

Table 1 summarizes the specifications of the instruments.

Table 1

| Spectrometer | 4G4K | 2G64K |
|---|---|---|
| RF input | 4 | 2 |
| ADC sample speed | 4.096 GHz | 2.048 GHz |
| Observation bandwidth | 2.048 GHz | 1.024 GHz |
| ADC resolution | 10 bit | |
| ENOB | 8 bit | |
| GFDR | ~38 dB | ~37 dB |
| Max. input power | 0 dBm | |
| Max. input frequency | ~8 GHz | |
| FPGA | Xilinx Virtex-7 (XC7VX485T) | |
| No. of FPGA per RF input | 1 | 2 |
| Frequency channel | 2048 | 32768 |
| Channel separation | 1 MHz | 31.25 kHz |
| Spectral resolution (−3 dB) | 2.0 MHz | 64 kHz |
| Spectral resolution (−10 dB) | 2.9 MHz | 90 kHz |
| Acquisition time per spectrum | 1 µs | 32 µs |
| Spectral dump time | 8 ms | |
| Number of accumulation | 8000 | 250 |
| FFT resolution | 32 bit | |
| Accumulator resolution | 45 bit | 40 bit |
| Output resolution | 16 bit | |
| Temperature range | 0–90 °C | |
| Allan variance | > 1500 s | |

*Test observations*

Our spectrometers have been installed in an 8-m parabola dish solar telescope at the National Institute of Information and Communications Technology (NICT). The observation bandwidth of this telescope is between 70 MHz and 9 GHz. Two 2G64K and eight 4G4K cover the entire observation band with both left and right circular polarizations simultaneously. The high frequency response of the spectrometers enables the direct sampling in the micrometric range and reduces the heterodyne frequency conversions in the receiver system. The observations started on August 2015. After that, the spectrometers have stably recorded the solar radio spectra every 8 ms. Figure 7 shows an example of the observed the solar radio burst. The top panel shows a solar radio type-II burst and middle panel shows a zoomed up image of the burst element indicated by the white rectangle in the top panel. Fine spectral structures with durations shorter than 1 s are recognized in the radio dynamic spectra. The bottom panel shows the radio burst observed at the same time and frequency range, by a traditional telescope HiRAS (Kondo et al, 1992). This telescope contains frequency-sweep type spectrum analyzers with the time and frequency resolutions of 500 ms and 500 kHz, respectively. Although the middle and bottom panels show similar spectral structures, only fine spectral structures with durations shorter than 1 s are recognized in the middle panel.

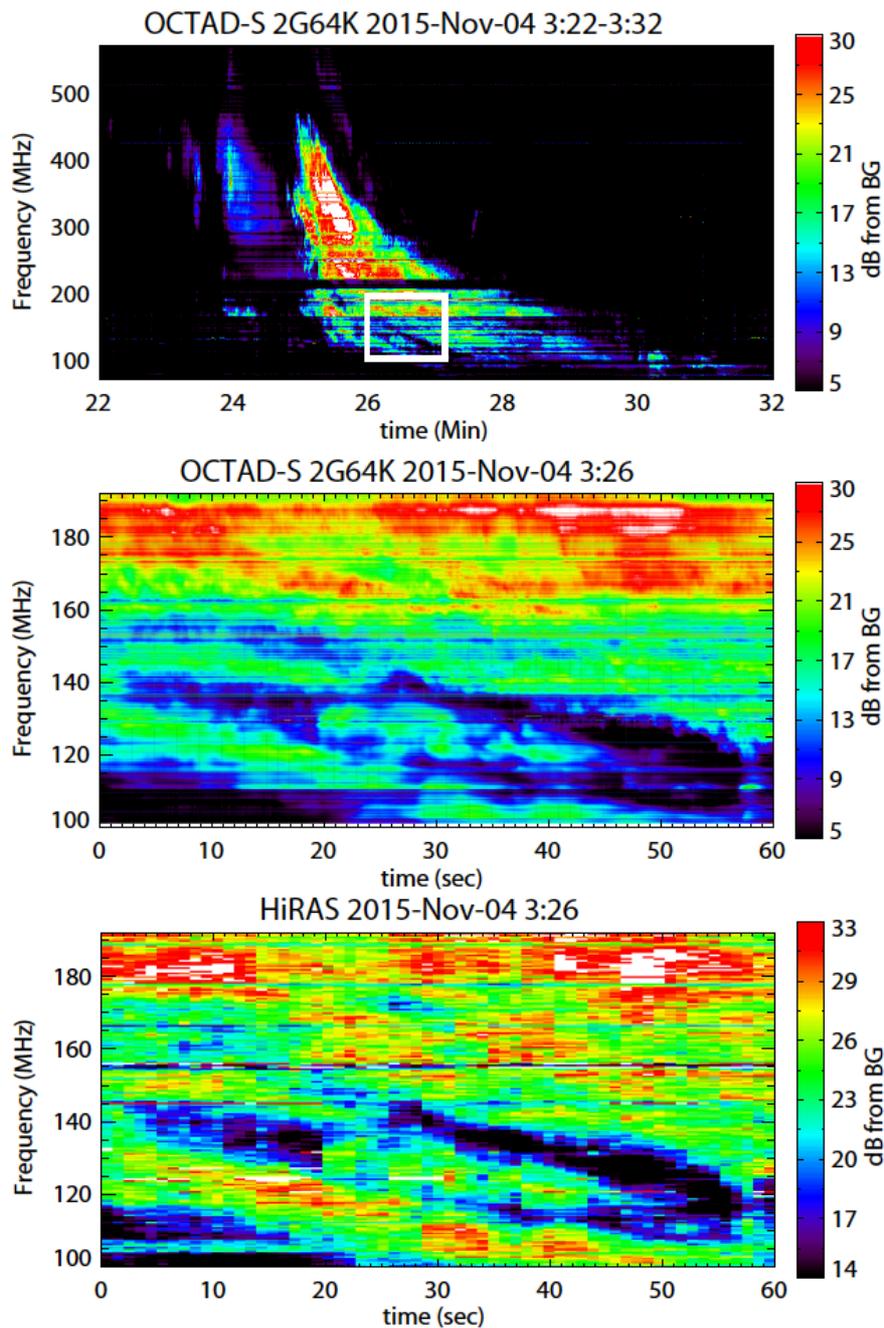

Figure 7. (Top) Solar radio type-II burst observed by OCTAD-S (2G64K) on Nov 4, 2015. (Middle) zoomed up image of one burst element indicated by the white rectangle in the top panel (observed between 3:26–27 UT). (Bottom) the solar radio burst observed by HiRAS at the same time and frequency ranges as the middle panel. Color of the radio dynamic spectra indicates the radio enhancement in dB from the background (BG). Horizontal lines in the radio dynamic spectra are artificial radio frequency interferences (RFIs).

## Conclusions

We developed a new digital FFT spectrometer named OCTAD-S using FPGA. The specifications and test results of the instruments are summarized as follows,

- The base instrument has independent ADC and FPGA modules, which allows us to change these devices.
- Two types of spectrometer have been implemented on the base instrument; one has a 4.096 GS/s sampling speed with 2048 frequency channels, the other one has a 2.048 GS/s sampling speed with 32768 frequency channels.
- A direct sampling spectroscopy up to 8 GHz is achieved by the microwave track-and-hold circuit, which can reduce the analog receiver in front of the spectrometer.
- The signal processing has no dead time and the accumulated spectra are recorded in the external media every 8 ms.
- Highly stable spectroscopy with the wide dynamic range of a 10-bit ADC was demonstrated in a series of experiments.

The dynamic range of the 10-bit ADC meets most radio-astronomical needs. However, a strong and wideband emission such as a solar radio burst sometimes requires a more dynamic range. Moreover, the wideband spectroscopy in the low frequency range always suffers from strong artificial RFIs, which are much stronger than the natural radio sources. A 12-bit (or more) ADC should be a more suitable solution. In addition, OCTAD-S used interleaving ADCs, which creates ghost spectra. A monolithic ADC would be better able to reduce the ghost spectra and increase the GFDR.

Wider band spectroscopy with more frequency channels is a general requirement for spectrometers. Increasing the frequency channels requires more signal processing resources in FPGAs. Improvement of the signal-processing algorithm will be essential for future digital spectrometers. The parallel FFT processing in FPGAs (Nakahara et al. 2013; 2014; 2015) would improve future digital spectrometers.

## List of abbreviations

ADC: Analog to digital converter

AOS: Acousto-optical spectrometers

DSP: digital signal processing

ENBW: equivalent noise bandwidth

ENOB: effective number of bits

FFT: Fast Fourier Transform

FPGA: Field-programmable gate array

GFDR: Ghost free dynamic range

FIR: Finite impulse response

OCTAD: Optically Connected Transmission system for Analog to Digital Conversion

PFB: Polyphase Filterbank

PLO: Phase-locked oscillator

RFI: Radio frequency interference

SFDR: Spurious free dynamic range

## Competing interests

The authors declare that they have no competing interests.

## Authors' contributions

KI led the performance evaluation experiments and drafted the manuscript. YK managed this project. HI and TN participated in the performance evaluation experiments. KH, KE, YH, YC developed and manufactured the instrument. All authors read and approved the final manuscript.

## Acknowledgements

KI is supported by a Japan Society for the Promotion of Science (JSPS) Research Fellowship. This project is partially supported by JSPS KAKENHI Grant Number 16K17813.